\documentclass[fleqn,usenatbib,useAMS]{mnras}

\usepackage{newtxtext,newtxmath}

\usepackage[T1]{fontenc}
\usepackage{ae,aecompl}

\usepackage{graphicx}	
\usepackage{amsmath}	
\usepackage{comment}

\newcommand{\ihmpc}{\,$h$\,Mpc$^{-1}$}
\newcommand{\hkpc}{\,$h^{-1}$\,kpc}
\newcommand{\hmpc}{\,$h^{-1}$\,Mpc}
\newcommand{\hgpc}{\,$h^{-1}$\,Gpc}
\newcommand{\hmsun}{\,$h^{-1}\,\mathrm{M}_{\odot}$}

\title[The Indra Simulations]{Indra: a Public Computationally Accessible Suite of Cosmological $N$-body Simulations}

\author[Falck, et al.]{Bridget Falck,$^1$\thanks{E-mail:bridget.falck@jhu.edu} 
	Jie Wang,$^{2,3}$ Adrian Jenkins,$^4$ Gerard Lemson,$^1$ Dmitry Medvedev,$^1$ \newauthor
	Mark C. Neyrinck,$^{5,6}$  Alex S. Szalay$^1$  \\
	$^1$Department of Physics and Astronomy, Johns Hopkins University, 3400 N Charles St, Baltimore, MD 21218, USA\\
    $^2$National Astronomical Observatories, Chinese Academy of Sciences, Beijing, 100012, China\\
    $^3$School of Astronomy and Space Science, University of Chinese Academy of Sciences, Beijing 100049, China\\
    $^4$Institute for Computational Cosmology, Department of Physics, Durham University, Durham DH1 3LE, UK \\
	$^5$Ikerbasque, the Basque Foundation for Science, 48009, Bilbao, Spain \\
	$^6$Department of Physics, University of the  Basque Country UPV/EHU, 48080, Bilbao,  Spain \\
}

\date{Accepted XXX. Received YYY; in original form ZZZ; Draft version \today}

\pubyear{2021}

\begin{document}
\label{firstpage}
\pagerange{\pageref{firstpage}--\pageref{lastpage}}
\maketitle

\begin{abstract}
Indra is a suite of large-volume cosmological $N$-body simulations with the goal of providing excellent statistics of the large-scale features of the distribution of dark matter. Each of the 384 simulations is computed with the same cosmological parameters and different initial phases, with 1024$^3$ dark matter particles in a box of length 1 \hgpc, 64 snapshots of particle data and halo catalogs, and 505 time steps of the Fourier modes of the density field, amounting to almost a petabyte of data. All of the Indra data are immediately available for analysis via the SciServer science platform, which provides interactive and batch computing modes, personal data storage, and other hosted data sets such as the Millennium simulations and many astronomical surveys. We present the Indra simulations, describe the data products and how to access them, and measure ensemble averages, variances, and covariances of the matter power spectrum, the matter correlation function, and the halo mass function to demonstrate the types of computations that Indra enables. We hope that Indra will be both a resource for large-scale structure research and a demonstration of how to make very large datasets public and computationally-accessible.
\end{abstract}




\section{Introduction}

Understanding the observations made by large-scale structure surveys such as Euclid, Roman, DESI, and the VRO requires input from theoretical predictions of structure formation in the form of numerical simulations. Cosmological $N$-body simulations solve for the nonlinear gravitational collapse of matter and are indispensable tools both for testing cosmological models and for planning for, and analyzing, observations. For example, simulations are necessary to predict the large-scale structure observables of dark energy and modified gravity theories in the non-linear regime~\citep[e.g][]{Joyce2016,Winther2015}, which can be compared to simulations of the current $\Lambda$CDM (cold dark matter with a cosmological constant, $\Lambda$) cosmological paradigm. Simulations have also become an integral part of the analysis of measurements of baryon acoustic oscillations (BAO) from galaxy redshift surveys~\citep[e.g.][]{Percival2014}.

One of the pressing challenges when making inferences from large observational surveys is that we can only observe the Universe from one vantage point in a finite volume. This ``cosmic variance'' places limits on the precision with which we can measure statistical fluctuations on very large scales. This also presents a problem for simulations: we cannot simulate exactly the positions of the galaxies in this observational volume. We can, however, simulate an ensemble of universes, increasing the precision with which we can numerically predict the statistical properties (e.g., the covariance matrix) of the matter distribution on large scales for a given cosmological model. 

Ideally, one would want to simulate the full volume and resolution that the next generation of galaxy surveys will attain, repeat this thousands of times for different realizations of the same cosmological model and parameters, run this ensemble for thousands of combinations of cosmological parameters in a given model, and repeat the entire process for any number of cosmological models one expects the observations to test. Such a scenario has obvious drawbacks: it would require a prohibitively expensive amount of computing power and produce an excessive amount of data which would be impossible to analyze without equally massive computing resources. Because of these immense technical challenges, workarounds are being devised: cosmological emulators~\citep{Heitmann2014,Heitmann2016,Garrison2018,DeRose2019,Nishimichi2019} attempt to span cosmological parameter space with a few full simulations and interpolate between them; low-resolution particle mesh simulations attempt to capture features only at large scales~\citep{Takahashi2009,Blot2015}; approximate simulations combine numerical shortcuts with analytic solutions to produce fast and cheap ensembles of realizations with a limited range of accuracy~\citep[see][and references therein]{Chuang2015,Lippich2019}; and analytic methods attempt to model the statistics in order to reduce the number of full realizations required to obtain the same precision~\citep{Pope2008,Taylor2013,Dodelson2013,Heavens2017}.

This paper aims to address the technical challenges head-on by producing a suite of cosmological $N$-body simulations called Indra\footnote{``Indra'' refers to the Buddhist metaphor of Indra's net, in which each of the infinite jewels reflects every other jewel.} hosted on SciServer\footnote{\url{http://www.sciserver.org}}, a science platform that allows the community to perform analysis where the data are stored. This method of \textit{server-side analysis} was pioneered in astronomy by the Sloan Digital Sky Survey (SDSS) SkyServer~\citep{Szalay2000}, which enabled astronomers to interactively explore a multi-terabyte database by leveraging state-of-the-art archiving technologies. Modelled on the success of the SkyServer, the Millennium Simulation Database~\citep{Lemson2006} made available the post-processing outputs of the Millennium simulation~\citep{Springel2005Mill}, including halo catalogs, merger trees, and later, semi-analytic galaxy catalogs~\citep[see also][]{Riebe2011,Loebman2014,Bernyk2016}. The dark matter particle positions and velocities, however, were not in the database, as their large size (roughly 20 TB) and usage patterns (being point clouds instead of object catalogs) presented significant technical challenges. 

Though database technology has seen significant advances in the past decade, our early efforts to build a relational database for Indra were motivated by reducing the number of rows in a table -- the planned full suite of 512 simulations amounts to 35 trillion output particles. We considered storing particles in spatially-sorted array-like ``chunks''~\citep{Dobos2012} and developed an inverted indexing scheme~\citep{Crankshaw2013} to access individual particles in these arrays (e.g. to track their movement across many snapshots). 
However, many technical advances opened up the capability to release data in their native binary file format, without needing to first load them into relational databases, by mounting the data to contained compute environments running on virtual machines. Indra takes advantage of the SciServer science platform, built at Johns Hopkins University where the data are stored, which provides interactive and batch mode server-side analysis in various environments, public and private file storage in both flat files and databases, and many collaboration tools~\citep{SciServer2020}. 

Recent advances have seen some tera- and peta-scale cosmological simulation suites that provide their data for download, but with no server-side computation~\citep{Blot2015,Quijote2020,Heitmann2019HACC}. Additionally, the latest version of the Illustris hydrodynamical simulations~\citep{Nelson2019} provides both data for download and an interactive compute environment similar to SciServer. Modern astronomical surveys are developing infrastructures and tools for the analysis and cross-matching of datasets that are hundreds of terabytes to petabytes in size~\citep{Juric2017}. The server-side analysis of both astronomical surveys and the theoretical predictions from simulations will soon become common practice for science involving very large datasets.

The Indra simulations are motivated by the goal of obtaining excellent statistics of the large-scale matter density field on 100\hmpc\ scales. We designed a suite of 512 (of which 384 are or will be currently available) realizations of 1\hgpc-sided cosmological $N$-body simulations, each with $1024^3$ dark matter particles. We saved 64 snapshots per simulation, many more than other simulation suites, so that ensemble statistics may be calculated for a large number of redshifts and merger histories of dark matter halos are able to be built. We describe the simulations and their data products in Section~\ref{sec:sims}, including details on how the initial conditions were created and an issue that affected the first 128 simulations. In Section~\ref{sec:results}, we measure ensemble averages, variances, and covariances from the three Indra data products: the dark matter particles, the halo catalogs, and the Fourier modes of the density field. In the appendix we describe how the full suite of simulations can be accessed and analyzed by the community.


\section{The Indra Simulations}
\label{sec:sims}

\subsection{Overview}

The Indra suite of simulations were run using the code L-Gadget2~\citep{Springel2005}, each with a box length of 1\hgpc, $1024^3$ particles, and WMAP7 cosmological parameters ($\Omega_M = 0.272$, $\Omega_\Lambda = 0.728$, $h = 0.704$, $\sigma_8 = 0.81$, $n_s = 0.967$)~\citep{Komatsu2011}. They are purely gravitational dark-matter-only cosmological simulations; each dark matter particle has a mass of $m_p = 7.031\times 10^{10}$\hmsun. The force softening length is 40\hkpc, and the parameters used for the force accuracy, time integration, and tree criteria are the same as used for the Millennium Run~\citep{Springel2005Mill}, which used the same lean version of Gadget. Initial conditions were generated using second-order Lagrangian perturbation theory~\citep[2LPT,][]{Scoccimarro1998,Jenkins2010}, with pre-initial conditions on a regular grid and a starting redshift of $z=127$.\footnote{ Since the inception of this project over a decade ago, there has been much progress in understanding the scale- and resolution-dependent effects of initial conditions generation and starting redshift in cosmological simulations. In particular, \citet{Michaux2021} quantify the discreteness effects of the initial lattice in simulations with the same box size and resolution as Indra. The difference between their reference ``FCC'' simulation and Indra's choice of $z_{init}=127$ with 2LPT varies with redshift, scale, and choice of statistic: for the power spectrum (their Figure 7), from a few percent at $z=3$ to 1\% at $z=0$ and $k=1$\ihmpc.} 64 snapshots are saved down to $z=0$ at the same redshifts as the Millennium Run. In general the box size, particle number, and other parameters of Indra are geared toward resolving halos more massive than that of the Milky Way and obtaining good statistics on BAO scales, while saving snapshots at a large number of redshifts and not exceeding a petabyte of data. Since Indra is intended to be a public resource, we hope it can serve a variety of cosmological applications.

The Indra simulations can be identified using three digits that range from 0 to 7, which we denote as x\_y\_z, corresponding to the coordinates of an $8\times 8\times 8$ cube. Note however that each simulation is fully independent and obeys periodic boundary conditions in a 1 (\hgpc)$^3$ volume; see Section~\ref{sec:ics} for details about how the initial phases were set up. Due to a bug in the initial conditions of the first 128 simulations, denoted 0\_y\_z and 1\_y\_z, these are not available, though we may re-run them in the future.

Halo catalogs were generated on-the-fly using a standard friends-of-friends algorithm~\citep[FOF;][]{Davis1985}, with a linking length of $b=0.2$ times the inter-particle separation, $L/N$, and a minimum of 20 particles per halo. After the simulations were finished, we ran SUBFIND~\citep{Springel2001} to identify subhalos in phase-space and calculate halo properties such as $R_{200}$ and $M_{200}$. 

The Indra simulations may be accessed via SciServer, a science platform that allows users to perform analysis where the data are stored without the need to download terabytes to a local compute cluster~\citep{SciServer2020}. Further details of data access are given in Appendix~\ref{sec:data_appendix}. Below we describe the three types of data output for each of the Indra simulations: 64 snapshots of particle positions and velocities, 64 snapshots of halo catalogs, and 505 snapshots of the Fourier modes of the density field. We then describe in detail the initial conditions of the simulations, which were created using Panphasia~\citep{Jenkins2013} to allow higher-resolution re-simulations of interesting regions.

At the time of this writing, we are finishing up the last series of 64 simulation runs and post-processing, and we will make these available as they are ready. Therefore, the analysis performed in this work is done on 320 Indra simulations: 2\_0\_0 to 6\_7\_7.

\subsection{Data Products}
\label{sec:data}

The output of the Indra simulations consists of three types of data products: the positions, velocities, and identifiers (IDs) of dark matter particles; halo and subhalo catalogs, along with the IDs of their constituent particles; and the Fourier modes of the coarse-gridded density field. The data are permanently stored on the DataScope, developed by the Institute for Data Intensive Engineering and Science\footnote{\url{http://idies.jhu.edu/what-we-offer/sciserver/datascope/}} at the Johns Hopkins University. The DataScope is a multi-petabyte file server with high throughput capability that hosts large datasets across many domains; Indra data are accessed from SciServer compute containers over a fast network connection. A subset of the data, including all halo catalog files and FFT outputs and some of the particle snapshots, is also stored on a distributed file system called the FileDB, which allows for more computationally-heavy analysis than the default SciServer Compute environment. Read times are somewhat faster for the data on FileDB, and a system for parallel computation using Dask\footnote{\url{http://dask.org/}} is in development. All data are accessed through SciServer using the indra-tools software package.\footnote{\url{http://github.com/bfalck/indra-tools}}  Both interactive Compute and batch Jobs run Docker containers on virtual machines, and the indra-tools software, whether run from a python script or Jupyter notebook, hides the filesystem from the user, choosing to read from FileDB if the requested data are there and from DataScope if not. Instructions on how to access the full suite of Indra data, as well as a description of the example Jupyter notebooks included with indra-tools, are provided in Appendix~\ref{sec:data_appendix}.

\subsubsection{Dark Matter Particles}

The bulk of the petabyte of Indra data are in the form of particle positions, velocities, and IDs for 64 snapshots of each simulation. These are stored as binary files in the standard Gadget format. The entirety of the particle snapshots are stored on the DataScope, and a subset of the particle data, consisting of a few full simulations and several full sets of snapshots, are also stored on FileDB servers. This subset might be slightly faster to work with and currently includes 5 full simulations (2\_0\_0, 3\_0\_0, 4\_0\_0, 5\_0\_0, and 6\_0\_0) and 9 full sets of snapshots at redshifts 0, 0.1, 0.5, 1, 1.4, 1.7, 2.0, 2.4, and 127.

\begin{figure}
	\includegraphics[width=\columnwidth]{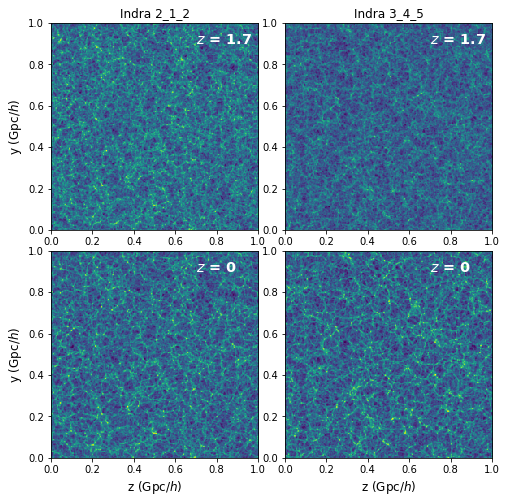}
    \caption{Slices of the logarithmically-scaled density field from 2 of the Indra simulations at 2 of the 64 snapshots, $z = 1.7$ and $z=0$, measured from CIC-interpolations of the $1024^3$ particle positions. Slices are 1 cell-length thick with 3.9\hmpc\ grid cells.}
    \label{fig:density}
\end{figure}

The snapshots of particle data are immediately available for analysis in the SciServer Compute environment. As an example, we present a visualization of the particle data in Figure~\ref{fig:density}, which shows the density field of slices through two of the 512 Indra simulations at $z=1.7$ and $z=0$. The density fields were measured from interpolations of particle positions onto a grid using the cloud-in-cell (CIC) assignment scheme. One CIC grid of a $1024^3$-particle snapshot can be computed in a few minutes using the package pmesh~\citep{pmesh}. We discuss preliminary analysis of the particle data, including power spectrum and correlation function covariances, in Section~\ref{sec:particle_results}.

\subsubsection{Dark Matter Halos}

The halo data consist of FOF group tables computed on-the-fly, SUBFIND halo catalogs computed after the runs have finished, and the IDs of the constituent particles of both the FOF and SUBFIND halos. The FOF data are minimal and contain only the number of particles in each halo, along with indexing information with which to pick out the correct particle IDs from the group ID data. Post-processing of the halos, including calculation of the mass, radius, position, velocity, and other halo properties is performed by SUBFIND and stored in the subhalo tables. SUBFIND identifies one subhalo as the ``main'' subhalo of that parent FOF group and measures quantities such as mass and radius for these halos, so in this way the subhalo catalog contains properties both of parent FOF groups and of their subhalos. 

For every FOF group, SUBFIND measures the radius within which the FOF group has an overdensity 200 times the \textit{mean} density of the simulation, the mass within this $R_{200,\mathrm{mean}}$, the radius within which the FOF group has an overdensity 200 times the \textit{critical} density of the simulation, the mass within this $R_{200,\mathrm{crit}}$, the radius within which the FOF group has an overdensity corresponding to the value at virialization in the top-hat collapse model for this cosmology, and the mass within this $R_{\mathrm{tophat}}$. For every subhalo, including the ``main'' subhalo, SUBFIND measures the 3-D coordinates of position as determined by the particle with the minimum gravitational potential, the 3-D coordinates of peculiar velocity, the 1-D velocity dispersion, the maximum of the circular velocity curve, the 3-D coordinates of halo spin, and the radius containing half of the mass. In the binary catalog data, all mass units are in \hmsun, radius and position units in \hmpc, velocity units in km/s, and spin units in (\hmpc)(km/s). In the database the units are the same except the masses are in units of $10^{10}$\hmsun.

The FOF and SUBFIND catalogs of the full suite of Indra simulations, for every snapshot, are stored in their binary file format in both the DataScope and FileDB filesystems. Additionally, they are loaded in to relational database tables, which can be queried directly from compute containers on SciServer. The usefulness of hosting halo catalogs in a relational database was successfully demonstrated by the Millennium Simulation database~\citep{Lemson2006}. They provide the capability of speeding up analysis in many cases, for example, selecting halos by mass from many snapshots and hundreds of Indra volumes without the need to read all of the binary catalog files. The Indra database also exploits a spatial indexing library developed for numerical simulations~\citep{LemsonDBLP} that enables efficient selection of halos or particles within shapes such as spheres or cones.

The suite of Indra subhalo data, saved both in binary files and in relational database tables, is a data-rich resource for studies of large-scale structure. For example, having 64 snapshots for each simulation allows the construction of halo merger histories, which isn't possible when only a few snapshots are stored. In addition, since all particle positions and velocities are saved, any other halo finder may be run to complement the existing FOF and SUBFIND catalogs, and new halo properties may also be measured from the constituent particles. We discuss preliminary analysis of the halo data, including the covariance of the mass function, in Section~\ref{sec:halo_results}.

\subsubsection{Fourier Modes of the Density}

The Fourier-space density field is output as the simulation runs, more frequently than are the particle snapshots, for a total of 505 time steps. The real and complex modes are defined on a $129 \times 129 \times 65$ grid of $(k_x,k_y,k_z)$, where $k_x$ and $k_y$ range from 0 to $\pm 0.4$\ihmpc\ and $k_z \geq 0$, which is sometimes denoted the ``upper-half-sphere'' of $k$-space. 
The largest wavenumber is thus $|\mathbf{k}|_{\mathrm{max}}=0.7$\ihmpc\ and correspondingly, the smallest length scale in the FFT data is 9 \hmpc. All of the FFT data, for every run and all 505 time steps, are stored on both the DataScope and FileDB filesystems and amount to about 2 TB. In Section~\ref{sec:fft_results}, we measure the distribution functions of the Fourier amplitudes of the density field and of the mode-dependent growth function, $D(a,k)$.

\subsection{Initial Conditions}
\label{sec:ics}

The initial conditions are generated using the IC\_2lpt\_Gen code~\citep{Jenkins2010}. Each Indra volume is a realization of a LCDM universe with cubic periodic boundary conditions.  To make the realizations independent and representative of the set of all possible volumes, we set the phases using the Panphasia multi-scale realization of a Gaussian white noise field~\citep{Jenkins2013,JenkinsBooth2013}.\footnote{\url{http://icc.dur.ac.uk/Panphasia.php}} Each Indra volume is assigned its own small cubic white noise field patch within the very large Panphasia field.  This individual white noise field is then convolved in $k$-space with an appropriate real non-negative filter to produce a realization of the LCDM power spectrum. The so-called $k$-space ``corner modes'', i.e. the power at modes larger than the one-dimensional particle Nyquist frequency, are not set to zero~\citep{Falck2017}.

The Panphasia realization is completely defined on all scales, which means once the cubic patch in the Panphasia field has been chosen, the phases for that Indra volume are known on all scales below the box scale -- even down to scales that could only be accessed by future zoom simulations in these volumes.

For reference we will give the locations of the phase information within the Panphasia field in this section.  The Panphasia field has an octree structure, so to describe a cubic region it is necessary to define five numbers: the level in the octree, three coordinates for the location of one corner of the cube, and the side-length of that cube. \cite{Jenkins2013} defined a fixed format text descriptor to hold these five pieces of information, in addition to a sixth check digit, for error checking purposes only, and the name for the volume. This text descriptor is used as an input by the IC\_2lpt\_Gen code to set the phases for a particular periodic volume. The descriptor, together with the cosmological parameters, box size and the linear power spectrum, is all that is needed to set up the Indra volumes themselves. Any future zoom simulations of any region in any of the Indra volumes, down to the putative CDM free streaming scale as need be, use the Panphasia descriptor for the parent volume.

The phases of all of the Indra volumes can be given as a set of descriptors that are functions of the three-digit identifiers $x\_y\_z$:
\begin{equation}
\nonumber
\mathrm{
[Panph1,L15,(X,Y,Z),S75,CH-999,INDRA\_xyz]}
\end{equation}
where $x,y,z$ are octal integers, each in the range of 0-7, that specify an Indra volume. 
The symbols, $X, Y, Z$, give the locations of one
corner of each cubic patch, where
\begin{eqnarray}
\nonumber
X = 31248 + 100*x, \\
\nonumber
Y = 31376 + 100*y, \\
\nonumber
Z = 31504 + 100*z. \\
\end{eqnarray}
We have set the check digits always to be -999 (which instructs the IC\_2lpt\_Gen code not to do the error checking). So for example the descriptor for the Indra\_555 volume is:
\begin{equation}
\nonumber
\mathrm{
[Panph1,L15,(31748,31876,32004),S75,CH-999,INDRA\_555]}
\end{equation}

 Although the Indra volumes are all close together in the Panphasia field, they are disjoint, which for a Gaussian white noise field means they are completely independent. Because of this, they cannot be mapped directly to an (8~\hgpc)$^3$ volume; in principle it would be possible to set up a realization of a roughly 10.7~\hgpc-sided cube that includes all of the Indra volume phase information. While individual patches of such a simulation would resemble individual Indra volumes to some extent, there would be clear differences on the largest scales as the periodic constraints are very different. In addition, for individual Indra volumes, the value of the integral of the white noise field over the entire simulation volume is ignored, as all Indra volumes must have exactly the mean density of the universe. In a larger volume this information would be retained, contributing to the largest scale modes of the periodic volume.


\section{Ensemble Averages, Variances, and Covariances}
\label{sec:results}

One major motivation for creating the Indra simulations is to enable precise measurements of ensemble averages, variances, and covariances at very large scales from full, not approximate, cosmological $N$-body simulations. In this section, we demonstrate the types of computations that Indra enables on each of the three main data products. Section~\ref{sec:particle_results} measures the mean, variance, and covariance of the matter correlation function and power spectrum from the dark matter particle positions. Section~\ref{sec:halo_results} focuses on the mass function of the dark matter halos. In Section~\ref{sec:fft_results}, we study the Fourier modes of the density field and measure the distribution function, and its evolution, of the Fourier amplitudes as well as the mode-dependent growth function $D(a,k)$.

\subsection{Matter Power Spectrum and Correlation Function}
\label{sec:particle_results}

\begin{figure*}
	\includegraphics[width=1.7\columnwidth]{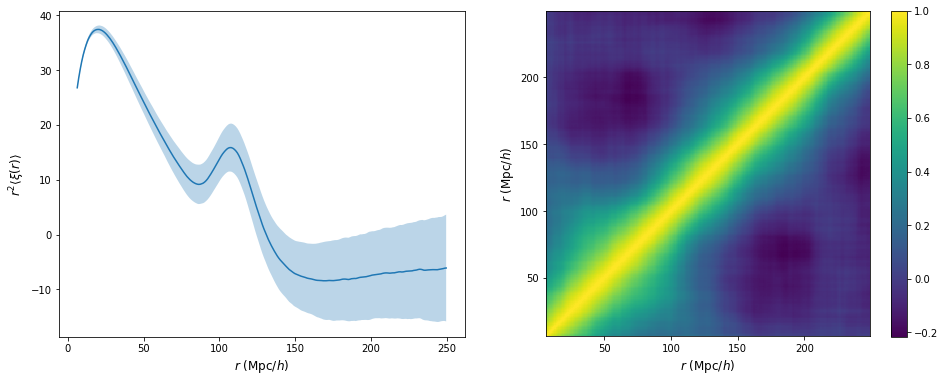}
    \caption{{\it Left:} The $z=0$ ensemble average correlation function measured from 320 Indra simulations. The shaded region indicates the standard deviation of the mean. {\it Right:} The correlation matrix of the same 320 correlation functions. 
    }
    \label{fig:xi_corr}
\end{figure*}

\begin{figure*}
	\includegraphics[width=1.7\columnwidth]{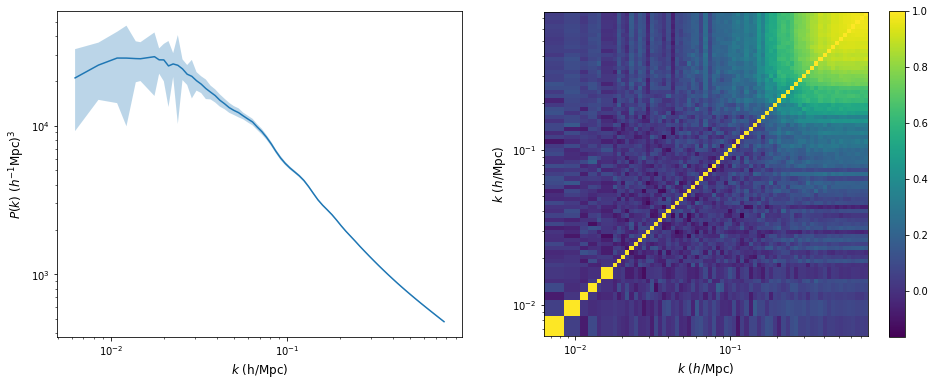}
    \caption{{\it Left:} The $z=0$ ensemble average power spectrum measured from 320 Indra simulations. The shaded region indicates the standard deviation of the mean. {\it Right:} The correlation matrix of the same 320 power spectra. }
    \label{fig:ps_corr}
\end{figure*}

The matter density contrast, $\delta = (\rho-\bar{\rho})/\bar{\rho}$, is a function of both space and time. Its spatial distribution can be summarized in terms of its two-point statistics in the form of the auto-correlation function, $\xi(\mathbf{r}) = \langle\delta(\mathbf{r}^\prime)\delta(\mathbf{r}^\prime+\mathbf{r})\rangle$, or its Fourier dual, the power spectrum, $P(k) \equiv \langle|\delta_k|^2\rangle$. These statistical descriptions completely specify a Gaussian random field, such as the density field in the very early Universe~\citep{BBKS1986}. The covariance matrix, $\mathbf{C}$, of a statistic $y$ is given by 
\begin{equation}
    C_{ij} \equiv \langle(y_i-\langle y_i\rangle)(y_j-\langle y_j\rangle)\rangle,
\end{equation}
where $i$ and $j$ are bins of either $r$ (for the correlation function covariance) or $k$ (for the power spectrum covariance).

Constraining cosmological parameters with large-scale structure surveys requires measuring the covariance matrix from a large number of mock galaxy catalogs~\citep[see, e.g.,][]{Percival2014,Lippich2019}, perhaps in combination with analytic methods~\citep[e.g.][]{Pope2008}. The error in covariance estimation due to the finite number of realizations propagates to errors in parameter constraints~\citep{Taylor2013,Dodelson2013,Sellentin2016,Blot2016}. Additionally, the mass resolution of the $N$-body simulation has a systematic affect on the covariance~\citep{Blot2015}. The distribution of power spectra departs from Gaussianity even at large scales~\citep{Takahashi2009,Blot2015}, necessitating the use of $N$-body simulations to study these effects.

We measure the correlation function and power spectrum of the Indra simulations by first interpolating the particle positions onto a grid using a cloud-in-cell (CIC) assignment scheme, which results in a measure of the density field in voxels of constant volume. This is then Fourier-transformed to obtain the Fourier modes of the density field, $\delta_k$, from which we compute the power spectrum and correlation function.

The correlation functions are measured from a CIC density grid of $1024^3$ cells (with a Nyquist frequency of 3.2\ihmpc). Figure~\ref{fig:xi_corr} shows the $z=0$ mean correlation function, and its variance, from 320 Indra simulations. The covariance of the correlation function is shown in the right panel of the figure in terms of its Pearson correlation matrix, 
\begin{equation}
    R_{ij} = \frac{C_{ij}}{\sqrt{C_{ii}C_{jj}}}.
\end{equation}

Figure~\ref{fig:ps_corr} shows the $z=0$ mean matter power spectrum and its variance, as well as the correlation matrix, of the dark matter particles of 320 Indra simulations, measured from a CIC density grid of $512^3$ cells (corresponding to a Nyquist frequency of 1.6\ihmpc). The CIC window function was deconvolved in the power spectrum calculation, but no shot-noise subtraction or anti-aliasing scheme was performed~\citep[see e.g.][]{Jing2005}.

Because we measure the correlation functions and power spectra over the full periodic simulation volumes, there are no modes present larger than our ``survey window'', so the covariance matrices do not contain the effects of beat-coupling~\citep{Hamilton2006} or local averaging~\citep{dePutter2012}. An observational galaxy survey would measure these effects in its covariance matrix, since modes are present in the Universe on scales larger than any survey volume. However, \citet{dePutter2012} found that these effects mostly cancel each other, and additionally for a (1 \hgpc)$^3$ volume, the correction to the power spectrum due to the local average effect is small, $\lesssim 10^{-4}$.

The Poisson error on the power spectra for 320 Indra simulations, given by $1/\sqrt{N(k)*320}$ for $N(k)$ modes at wavenumber $k$, is 2.5\% at the largest scales where $k\sim 0.01 $\ihmpc\ and 0.3\% at $0.06$\ihmpc, roughly the BAO scale. The Indra simulations provide an excellent $N$-body benchmark for detailed studies of the matter power spectrum and correlation function covariance using analytic and approximate methods.

\subsection{Halo Mass Function}
\label{sec:halo_results}

In the context of understanding or analyzing large-scale galaxy surveys, the halo catalogs are of primary importance. Methods which aim to speed up $N$-body simulations by introducing approximations or simplifications of the fully nonlinear gravitational collapse focus on how well they can reproduce the spatial distribution, masses, and key characteristics of the dark matter halos identified in full $N$-body simulations~\citep{Lippich2019,Blot2019}. In this data release paper, we focus on the mass distribution of the Indra halos and its ensemble average, variance, and covariance. Specifically, we measure the $M_{200}$ mass function of the main subhalos identified by SUBFIND, excluding the children of the main subhalos. $M_{200}$ is defined as the mass within $R_{200}$, the radius within which the density is greater than 200 times the critical density, $\rho_c$.

\begin{figure}
	\includegraphics[width=\columnwidth]{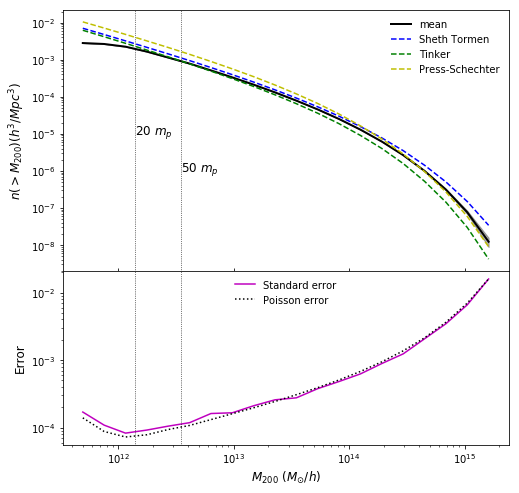}
    \caption{ {\it Upper:} Average $M_{200}$ mass function at $z=0$ of 320 simulations. Shaded gray region shows the standard deviation in each mass bin, and vertical lines show where the $M_{200}$ mass is equivalent to 20 and 50 times the particle mass. Theoretical mass functions from \citet{Sheth2001,Press1974,Tinker2010} are also shown. {\it Lower:} Standard deviation of the $z=0$ mass functions of Figure~\ref{fig:m200}, along with the Poisson error. The Poisson error is a very good approximation of the standard deviation of the mass function for larger halos, but underestimates it for smaller halos.}
    \label{fig:m200}
\end{figure}

The upper panel of Figure~\ref{fig:m200} shows the average $M_{200}$ mass function at $z=0$ of 320 Indra simulations, along with theoretical mass functions \citep{Sheth2001,Press1974,Tinker2010} derived using \mbox{HMFcalc}~\citep{Murray2013}.\footnote{\url{http://hmf.icrar.org/}} 
The standard deviation of these mass functions is given by the shaded grey area around the mean and is only visible for the highest masses. As is standard, the mass function begins to flatten at masses for which halos are not well-resolved due to the resolution limit of the simulation. Vertical lines show where the halo mass is equivalent to 20 and 50 times the particle mass, though note that these masses do not correspond to the number of particles in the halos (all halos have a minimum of 20 particles) since we have measured the $M_{200}$ mass function; the lines are merely a guide to suggest a point along the mass function at which halos may be considered well-resolved, which is a subjective judgment in any case. 

Though the mass function is generally plotted as the cumulative number density of halos greater than a given mass as a function of mass, the variance of the mass function is better characterized by the mean number of halos (not the cumulative number density) in each mass bin, $\bar{N}$. The bottom panel of Figure~\ref{fig:m200} shows the relative standard deviation of the mean number of halos in each mass bin, $(\sigma_{\bar{N}}/\bar{N})/\sqrt{320}$, along with the Poisson error, $1/\sqrt{\bar{N}*320}$. The Poisson error is a very good approximation to the standard deviation of the mass function for halos above $\sim 10^{13}$ \hmsun\ and slightly underestimates the standard deviation for lower mass halos. The errors generally increase as a function of mass, as the number of halos at a given mass decreases, but also increase at low mass, as fewer halos are found due to the resolution limit of the simulation.

\begin{figure}
    \includegraphics[width=\columnwidth]{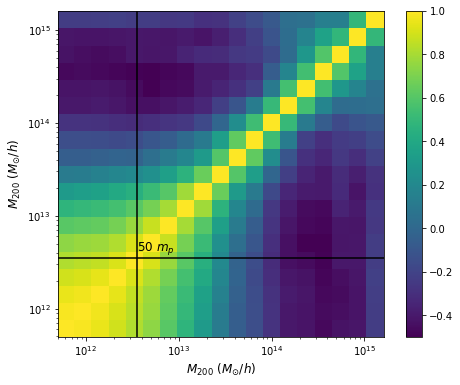}
    \caption{Correlation matrix of the mass functions of 320 Indra simulations at $z=0$. Horizontal and vertical lines show where the $M_{200}$ mass is equivalent to 50 times the particle mass.}
    \label{fig:m200corr}
\end{figure}

Figure~\ref{fig:m200corr} shows the correlation matrix of 320 $z=0$ mass functions. As a guide, the horizontal and vertical lines show where the $M_{200}$ mass is equivalent to 50 times the particle mass, or $3.5\times 10^{12}$\hmsun. The off-diagonal elements are greatest for the lowest mass halos, echoing the small-scale (large $k$) behavior of the power spectrum covariance in Figure~\ref{fig:ps_corr}. The covariance matrix has a rather rich structure that to our knowledge we are the first to measure to this precision. There is a substantial anti-correlation far from the diagonal, i.e. between large and small halos. We suspect this is a result of conservation of mass in these simulations, i.e., larger halos grow at the expense of smaller halos.

\subsection{Fourier Modes of the Density Field}
\label{sec:fft_results}

In this section we analyze the Fourier modes of the density field that are output as the simulations run. These data enable the study of the mode-by-mode evolution of the density field without the need to measure computationally-expensive density grids from the particle positions, though only down to scales of $k\sim 0.7$\ihmpc\ (9 \hmpc). We first measure the mean and variance of the $z=0$ power spectra and compare this to those derived from the particle positions. We then look in detail at the one-point distribution functions of the density, $\delta_k(a)$, and the mode-dependent growth function, $D(a,k)$, in four bins of wavenumber.

\subsubsection{Power Spectra}

\begin{figure}
\includegraphics[width=\columnwidth]{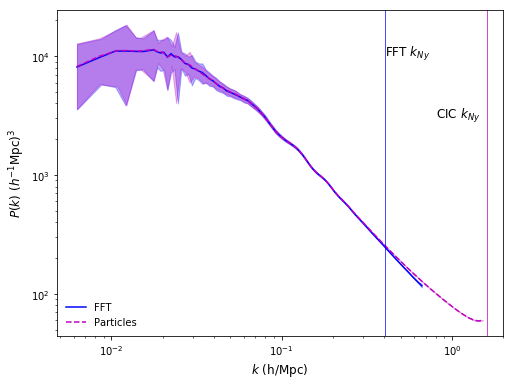}
\caption{Ensemble average power spectra of 320 simulations at $z=1$. In blue is the mean and standard deviation of power spectra derived from the FFT output, and in magenta is the mean and standard deviation of power spectra from CIC density fields of the particle positions (as in Section~\ref{sec:particle_results}). Vertical lines show the Nyquist frequencies of the FFT output (blue) and of the CIC grid (magenta).}
\label{fig:fftpk_z1}
\end{figure}

Figure~\ref{fig:fftpk_z1} shows the mean and variance of the power spectra of 320 Indra realizations at $z=1$, measured both from the output Fourier modes of the density field and from the CIC density grid derived from the particle positions. Both the mean and the variance of the FFT and particle-based power spectra agree very well with each other up to the Nyquist frequency of the FFT output, with only a slight deviation at higher frequencies (at the corner modes; see Section~\ref{sec:ics}). Note that the FFT and CIC power spectra were not computed in the same $k$-space bins, which explains some of the disagreement at low frequencies. The FFT outputs are therefore an exceptional resource for studies of the Fourier-space density field from the largest scales down to the mildly nonlinear regime of 10 \hmpc, at very high temporal resolution, without the need for expensive interpolations of the density field from the particle data. These data can also be an excellent training set for neural network architectures.

\subsubsection{One-point Distribution Function}

The properties of the distribution of density fluctuations at early times are well-described by a Gaussian random field, as extensively studied by~\citet{BBKS1986}. In particular, the Fourier transform of a Gaussian random field allows analytical study of the decomposition of the field into its Fourier amplitudes and phases. 
For a given scale-factor $a$ and wavenumber $k$, the Fourier amplitudes of the density field can vary for different random initial conditions and for different wave-vectors within a bin of wavenumber. We can thus measure the one-point distribution function of Fourier modes from an ensemble of simulations such as Indra and study its evolution and scale-dependence.

In \citet{Matsubara2007}, the complex Fourier modes are normalized by the power spectrum, and a distribution function is obtained for
\begin{equation}
    \alpha_k = \frac{\delta_k(a)}{\sqrt{P(a,k)}} = \frac{\delta_k(a)}
    {\sqrt{\langle|\delta_k(a)|^2\rangle}} = A_k\,e^{i\theta_k}.
\end{equation}
\citet{Matsubara2007} finds that the distribution function of a Fourier mode does not depend on the phase $\theta_k$, with the result that, for a random field (not necessarily Gaussian) in a spatially homogeneous space, the one-point distribution function of Fourier phase, $P(\theta_k)$, is always homogeneous, that is, the phases are uniformly distributed. 

\begin{figure}
\includegraphics[width=\columnwidth]{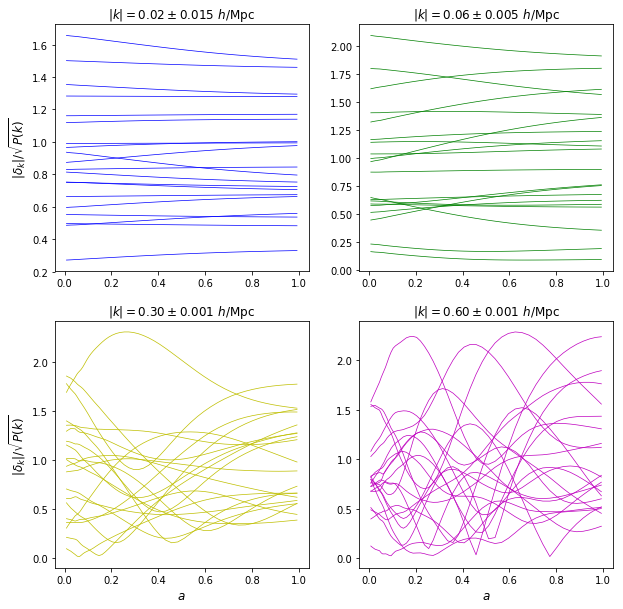}
\caption{Evolution of normalized Fourier modes of the density field for a random selection of wave-vectors in a bin of wavenumber in one Indra simulation.}
\label{fig:dk_evol}
\end{figure}

\begin{table}
    \centering
\begin{tabular}{c|c|c|c|c}
    $|\mathbf{k}|$ (\ihmpc) & $\delta k$ & $l$ (\hmpc) & $(l_{min}, l_{max})$ & $N_m$ \\
     \hline
    0.02 & 0.015 & 314 & (180, 1256) & 417 \\
    0.06 & 0.005 & 105 & (97, 114) & 998 \\
    0.30 & 0.001 & 20.9 & (20.87, 21.01) & 4548 \\
    0.60 & 0.001 & 10.47 & (10.45, 10.49) & 1224 \\
\end{tabular}
    \caption{The wavenumber bins, their corresponding range of scales, and the number of wavevectors (for one simulation) in each bin. Thus the total number of modes in each bin is $N_m$ multiplied by the number of simulations used, which is 320 in this paper.}
    \label{tab:kbins}
\end{table}

\begin{figure*}
\includegraphics[width=1.5\columnwidth]{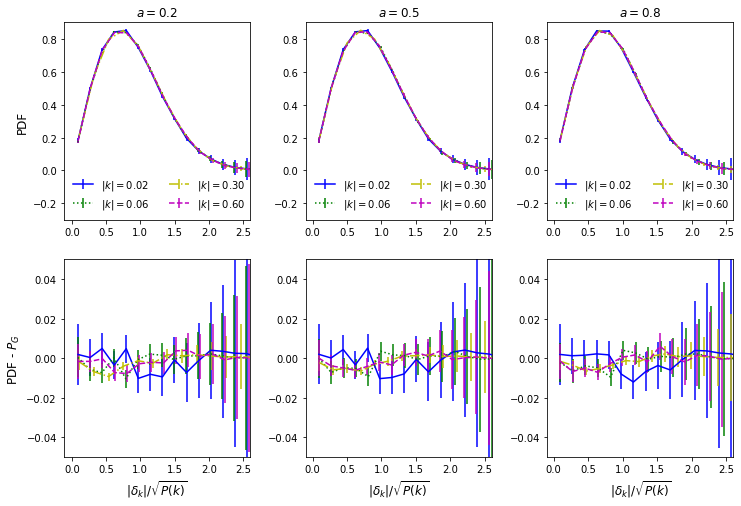}
\caption{{\it Upper panels:} One-point distribution functions of normalized Fourier amplitudes of the density field in four $k$ bins and at three values of the scale factor, measured for 320 Indra simulations. {\it Lower panels:} Difference between the Indra distribution functions and those of a Gaussian random field. Error bars correspond to Poisson errors.}
\label{fig:dk_hists}
\end{figure*}

\begin{figure}
\includegraphics[width=\columnwidth]{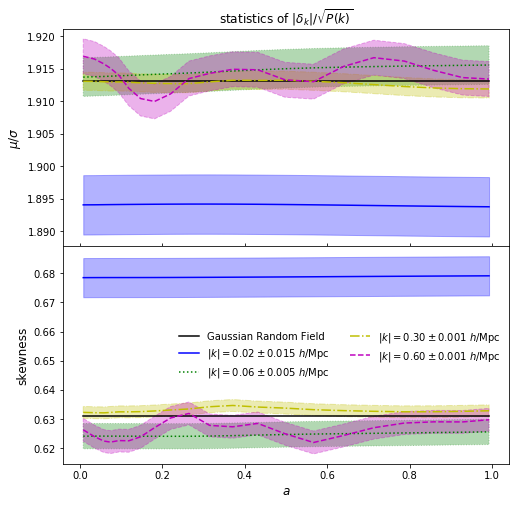}
\caption{Statistics of Fourier mode distribution functions as a function of scale-factor in four $k$ bins. The upper panel shows the ratio of the mean to the standard deviation, along with its value for a Rayleigh-distributed Gaussian random field, while the lower panel shows the skewness and its Gaussian value. The bands show the standard errors of each statistic.}
\label{fig:dk_stats}
\end{figure}

Figure~\ref{fig:dk_evol} shows the evolution of individual Fourier modes for 20 random wave-vectors having nearly the same wavenumber in four $k$ bins, defined in Table~\ref{tab:kbins}: the largest modes (upper left), modes around the BAO scale (upper right), the regime where $L\sim 21$\hmpc\ (lower left), and the mildly nonlinear regime where $L\sim 10.4$\hmpc\ (lower right). As $k$ increases and the length scale gets smaller, the evolution of a given Fourier mode becomes more nonlinear, but at all scales, the range of normalized mode amplitudes within the $k$ bin is similarly spread out. These trajectories look rather random, but are deterministic, driven by the evolving phase and amplitude couplings between each mode and all other modes (i.e.\ the rest of the density field) \citep[e.g.][]{BernardeauEtal2002}. The distribution of mode amplitudes at a given time is the one-point distribution function studied by ~\citet{BBKS1986} and~\citet{Matsubara2007}.

The Fourier modes of a Gaussian random field have a distribution function that is Rayleigh-distributed~\citep{BBKS1986}:
\begin{equation}
    P_\mathrm{G}(A_k,\theta_k)\,dA_kd\theta_k = 2A_k e^{-A_k^2}\,dA_k\frac{d\theta_k}{2\pi}
    \label{eqn:pgauss}
\end{equation}
(see Equation 68 in \citet{Matsubara2007}). In the upper panels of Figure~\ref{fig:dk_hists}, we show the one-point distribution function of Fourier modes from 320 Indra simulations for the four $k$ bins given in Table~\ref{tab:kbins} and three values of the scale factor, $a$ = 0.2, 0.5, and 0.8, corresponding to redshifts of $z$ = 4, 1, and 0.25. The lower panels of Figure~\ref{fig:dk_hists} show the difference between these distribution functions and that of a Gaussian random field. All of the distribution functions appear to be relatively consistent with a Rayleigh-distributed Gaussian random field. 

As Figure~\ref{fig:dk_hists} shows, there is very little difference between the distribution functions at different times and scales. However, Figure~\ref{fig:dk_evol} shows differences in the time-evolution of individual Fourier modes at different scales. In Figure~\ref{fig:dk_stats} we show summary statistics of the one-point distributions of Fourier modes as a function of scale factor. The upper panel shows the ratio of the mean to the standard deviation, $\mu/\sigma$, which has a constant value of $\sqrt{\pi/(4-\pi)} \simeq 1.913$ for a Rayleigh-distributed Gaussian random field. The lower panel shows the evolution of the skewness, which is equal to $(2\sqrt{\pi}(\pi-3))/(4-\pi)^{3/2} \simeq 0.631$ for a Gaussian random field. The error bands are given by the standard errors of each statistic, assuming that the statistics are Gaussian distributed~\citep{Harding2014}. The largest-scale mode, at $k = 0.02$\ihmpc, is the only one for which the statistics lie outside those of a Rayleigh distribution. This may be because there is a factor of 2 fewer modes in this largest bin than the $k = 0.06$\ihmpc\ bin. It is also interesting to note the oscillation of the distribution function statistics of the smallest-scale mode, $k = 0.6$\ihmpc, but further investigations are outside the scope of this paper.

\subsubsection{Scale-dependent Evolution}

In linear theory, the matter density contrast $\delta$ evolves independently of scale. The \textit{linear growth function} $D_L(a)$ can be defined via
\begin{equation}
    \delta(a) = D_L(a)\delta(a=1).
    \label{eqn:Dlin}
\end{equation}
For a flat universe where $\Omega_M+\Omega_\Lambda=1$,
\begin{equation}
    D_L(a)=5/2\,\Omega_M\sqrt{\Omega_M/a^3+\Omega_L} \int_0^1\,\frac{da}{\left(a\sqrt{\Omega_M/a^3+\Omega_L}\right)^3},
    \label{eqn:Dtheory}
\end{equation}
which reduces to $D_L(a)=a$ when $\Omega_M=1$.

When $\delta \gtrsim 1$, the evolution becomes nonlinear, and the $k$-modes no longer evolve independently, an effect known as mode-coupling~\citep{Meiksin1999,Scoccimarro1999}. In this case, one can solve for the evolution of the density field perturbatively:
\begin{equation}
    \delta(a,k) \approx D(a)\delta^{(1)}(k) + D^2(a)\delta^{(2)}(k)+ ...
\end{equation}
Taking the first order only, and approximating $\delta^{(1)}(k)$ as the density field of the first output of the simulations at $z=127$, we measure the \textit{mode-dependent growth function} $D(a,k)$ as the amplitude of the ratio of the complex Fourier modes of the density field, 
\begin{equation}
    D(a,k) = |\delta_k(a)/\delta_k(z=127)|.
    \label{eqn:D}
\end{equation}
We then divide this by the mode-independent linear growth function, calculated from the analytic solution in Equation~\ref{eqn:Dtheory} and scaled by the growth function at $z=127$:
\begin{equation}
    D(a) = D_L(a)/D_L(z=127).
\end{equation}
The ratio is then
\begin{equation}
    D(a,k)/D(a) = \frac{|\delta_k(a)/\delta_k(z=127)|}{D_L(a)/D_L(z=127)}.
    \label{eqn:Dratio}
\end{equation}
When this ratio deviates from unity, that is a sign that the evolution is nonlinear and the mode-coupling term is important.

\begin{figure}
	\includegraphics[width=\columnwidth]{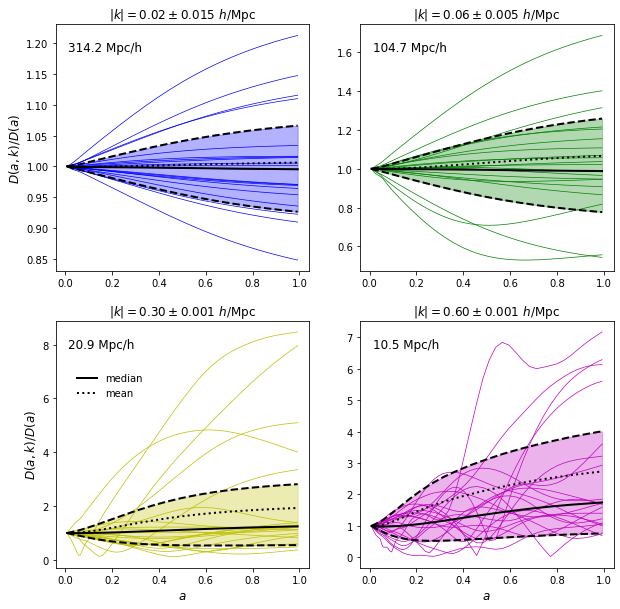}
    \caption{Mode-dependent growth function evolution for a random selection of wave-vectors having the same wavenumber, as well as the median (solid), mean (dotted), and 16th and 84th percentiles (dashed) of the distributions of all wave-vectors having within the given range of wavenumber over 320 Indra simulations. Wavelengths of modes in the middle of each bin appear in the upper-left corner of each panel.}
    \label{fig:mode_vecs}
\end{figure}

The collection of $D(a,k)$'s that contribute to a $k$ bin can exhibit a large amount of variation for one simulation and across many realizations of random initial phases. 
Figure~\ref{fig:mode_vecs} shows the evolution of the mode-dependent growth function for 20 random wave-vectors in the four $k$ bins defined in Table~\ref{tab:kbins}. Note that the y-axis ranges change for each wavenumber. Also shown are the median, mean, and inner 68\% percentiles of the distributions of all wave-vectors within each wave-number range over 320 Indra simulations, shown in black.

\begin{figure}
	\includegraphics[width=\columnwidth]{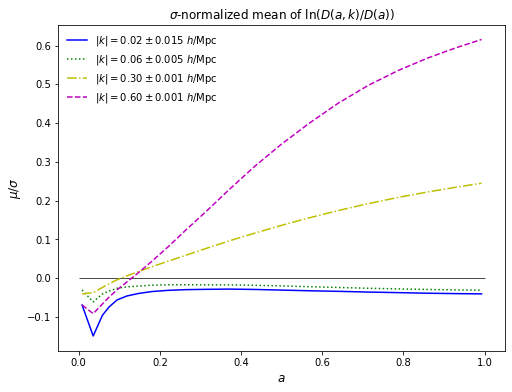}
    \caption{Evolution of the sigma-normalized mean of the logarithm of the mode-dependent growth function, $D(a,k)$.}
    \label{fig:lnD_stats}
\end{figure}

At a given $a$, the distribution of mode-dependent growth functions for the largest mode is very nearly Gaussian, while the BAO (105\hmpc), mildly nonlinear (20.9\hmpc), and more nonlinear (10.5\hmpc) distrubutions are positively skewed. It turns out that these distributions, for a given scale factor, are very nearly lognormal. In Figure~\ref{fig:lnD_stats} we plot the sigma-normalized mean of the logarithm of these distributions, or explicitly, $\langle\ln(D(a,k)/D(a))\rangle$, as a function of the scale factor. Here, the average $\langle\rangle$ is taken over the simulations in the Indra ensemble and the modes in each $k$ bin. Notably, the means of the two largest-scale bins are below zero, meaning the means of the mode-dependent growth functions are less than the linear growth rate, at all scale factors, while the two smaller-scale modes are above zero and grow over time. This shows how energy cascades from large to small scales as a result of mode-mode coupling.

\begin{figure*}
	\includegraphics[width=2\columnwidth]{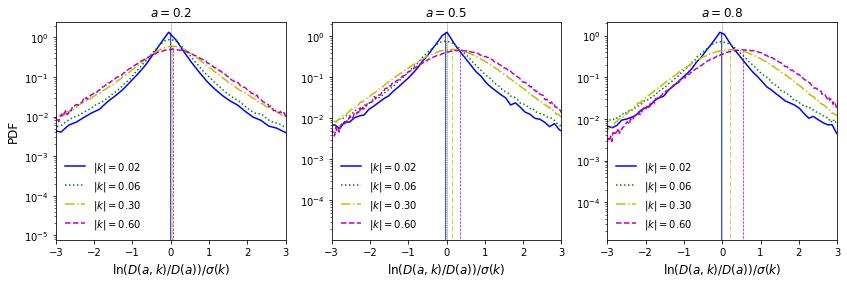}
    \caption{Distribution functions of $\sigma$-normalized $\ln(D(a,k)/D(a))$ in four $k$ bins and at three values of the scale factor, measured for 320 Indra simulations. The $\sigma$-normalized mean values of the distributions are given as vertical lines under the histograms, showing that the two larger scale modes have means less than 0, and the two smaller-scale modes have means greater than 0. See Fig.~\ref{fig:mode_vecs} for a sense of  distribution widths without $\sigma$-normalizing; the distributions are narrow at low $k$.
    }
    \label{fig:lnD_hists}
\end{figure*}

As with the Fourier mode distribution functions in Figure~\ref{fig:dk_hists}, it is illustrative to look at the full distributions of the mode-dependent growth function at a given scale-factor. 
In Figure~\ref{fig:lnD_hists} we show the distributions of $\ln(D(a,k)/D(a))$ in four $k$ bins, normalized by their standard deviations, for three different scale factors. Recall from Equation~\ref{eqn:D} that we have defined the mode-dependent growth function as the amplitude of the complex Fourier modes of the density field, $\delta_k(a)$, normalized by $\delta_k(z = 127)$ as a proxy for the linear density field, and that the ratio $D(a,k)/D(a)$ is near unity when the density field evolves linearly. 
Even after normalizing the width of the histograms by their standard deviations, which is much larger for the smaller-scale modes, the distributions become wider as $k$ increases and the growth function tends to deviate far from the linear theory value. The distribution of the largest-scale mode ($|k|$ = 0.02\ihmpc) is sharply peaked around 0, which is expected for linear evolution.


\section{Conclusion}
\label{sec:conc}

Indra is a suite of cosmological $N$-body simulations made available and computationally-accessible through the SciServer science platform. Each of the 384 (\hgpc)$^3$ volumes was run with the same cosmological parameters and different realizations of the initial conditions, providing the means to calculate precise averages and covariances on large scales. The large number of saved snapshots (64) for each simulation provides excellent redshift coverage and enables the calculation of halo merger trees, semi-analytic catalogs, and other derived data products.

In this paper, we have presented ensemble averages, variances, and covariances for each of Indra's three main data products: dark matter particles, halo catalogs, and Fourier-space density grids. We also studied in detail the one-point distribution function of the Fourier-space density field and the evolution and mode dependence of the growth function. We anticipate that the suite of Indra simulations will enable a variety of scientific investigations, such as:
\begin{itemize}
\item conditional statistics: mass functions, correlation functions, and power spectra conditional on e.g. a density criterion \citep[see, e.g.][]{Massara2020};
\item extreme statistics, e.g. of very large or very rare structures \citep[see, e.g.][]{Watson2014};
\item re-simulations of interesting structures at higher resolution (enabled by Panphasia~\citep{Jenkins2013} initial phases);
\item new databases of structures: halos from other halo finders, voids or filaments defined by halos or by  the dark matter particles, etc.;
\item ``event database'' of interesting things from the full suite, e.g. merging clusters;
\item suites of mock galaxy catalogs and lightcone catalogs, optimized by fast spatial searches and well-sampled over a large range of redshifts;
\item studies where a full $N$-body reference ensemble would benchmark tests of, or eliminate the need for, suites of approximate simulation methods; and
\item the development of machine learning methods requiring large training sets of structure formation on 10 to 100~\hmpc\ scales.
\end{itemize}

The ability to ask these and other scientific questions of the petabyte-scale Indra simulations is supported by the data storage and computational infrastructure of the DataScope and SciServer~\citep{SciServer2020}. 
For modern data sets of the size as the Indra suite, it becomes unfeasible to require users to download the data for local analysis. Instead, a science platform such as SciServer allows users to bring their analysis to the data. In this paper we have presented patterns of storage (database and distributed file systems) and compute (Jupyter notebooks in Docker containers) developed for SciServer that we found useful for large simulation data and that can in principle be replicated by other teams. SciServer itself has been deployed in a number of other institutes and various groups are building their own science platforms. Current and future large-scale structure surveys like Euclid, Roman, DESI, and the VRO also require large suites of simulations to perform their analysis, and we would strongly encourage those groups to publish any simulation data catalogues together with their observational data.

We believe that making Indra available in an environment such as SciServer is worth the considerable effort and resources. Making the data explorable and interactive, without the need to download terabytes to a local compute environment, opens up discovery space to the community of simulation experts as well as students and the public.

\section*{Acknowledgements}

We could not have made Indra publicly available without the support of IDIES and SciServer developers and engineers, including Brian Mohr, Jan Vandenberg, Manu Taghizadeh-Popp, Sue Werner, and Lance Joseph. We are thankful for the work done on earlier ideas for a public Indra database by Daniel Crankshaw, Tam\'{a}s Budav\'{a}ri, and L\'{a}szl\'{o} Dobos and for preliminary scientific analysis performed by Nuala McCullagh. We are grateful to Volker Springel for providing us with the L-Gadget2 and SUBFIND codes, and our analysis made extensive use of the following code packages: ipython~\citep{ipython}, matplotlib~\citep{matplotlib}, numpy~\citep{numpy}, and scipy (\url{http://www.scipy.org/}). This work could not have been carried out without support from the administrative and custodial staff at several institutions. MCN is grateful for funding from Basque Government grant IT956-16.

Computational resources for the Indra simulations were provided by the Homewood High-Performance Cluster (HHPC) and the Maryland Advanced Computing Center (MARCC) through the Institute for Data-Intensive Engineering and Science (IDIES) at JHU. This research used resources of the National Energy Research Scientific Computing Center (NERSC, ROR https://ror.org/05v3mvq14, GRID ID grid.484489.d), a U.S. Department of Energy Office of Science User Facility located at Lawrence Berkeley National Laboratory, operated under Contract No. DE-AC02-05CH11231. Work relating to the initial conditions used the DiRAC Data Centric system at Durham University, operated by the Institute for Computational Cosmology on behalf of the STFC DiRAC HPC Facility (\url{http://www.dirac.ac.uk}). This equipment was funded by BIS National E-infrastructure capital grant ST/K00042X/1, STFC capital grants ST/H008519/1 and ST/K00087X/1, STFC DiRAC Operations grant ST/K003267/1 and Durham University. DiRAC is part of the UK National E-Infrastructure.

\section*{Author Contributions}

BF did the bulk of the analysis and wrote the majority of this manuscript. BF coordinated the data storage, kept track of the simulation and post-processing progress, and worked with GL and DM to set up and test the public access framework on SciServer. BF wrote the python package that reads Indra data and the example notebooks that show how to interact with Indra on SciServer. JW maintained the production Gadget and SUBFIND codes, coordinated supercomputer time, and ran the bulk of the simulation and the post-processing codes. ARJ helped with planning and setting up the initial conditions and contributed to the writing of part of Section 2 of the manuscript. GL is the projects manager of SciServer and provided scripts for loading the databases and for the 3D spatial query libraries. DM developed critical infrastructure to enable public access and computational support of Indra on SciServer. MCN helped to plan the simulations and analysis, coordinated supercomputer time for the simulations, and ran several of them. AS helped to procure funding, plan the project, and analyze the results.

\section*{Data Availability}

The data underlying this article are available from SciServer at \url{http://sciserver.org}. 
The derived data generated in this research will be shared on reasonable request to the corresponding author.



\bibliographystyle{mnras}
\bibliography{indra}



\appendix

\section{Accessing and analyzing the Indra simulations}
\label{sec:data_appendix}

Indra is accessed by creating an account on the SciServer science platform~\citep{SciServer2020} at \url{http://www.sciserver.org}. SciServer users have access to public datasets and SQL databases, private data storage, file-sharing and group collaborative tools, and the interactive and batch-mode compute environments. Users effectively opt-in to datasets they are interested in by choosing the appropriate \textit{Science Domain}; Indra is in the \textit{Cosmological Simulations} Science Domain, which also contains data from the Millennium simulations. Once a user joins this Domain, they will be able to query the Indra database in \textit{CasJobs} and load the Indra data and software in a \textit{Compute} container, both of which are reached from SciServer \textit{Dashboard} after you log in. General SciServer tutorials can be found at \url{http://www.sciserver.org/support/help/}.

All of the Indra data, including the database of halo catalogs, can be analyzed in a \textit{Compute} container. When creating a new container, choose the \textit{Cosmological Simulations} compute image, which has indra-tools\footnote{\url{http://github.com/bfalck/indra-tools}} and some other simulation software pre-installed, and check all of the Indra Data Volumes. 
The indra-tools github repository includes several example Jupyter notebooks to help get users started with accessing the data and performing some analysis:
\begin{itemize}
    \item \textit{read\_examples}: How to read all of the data products: snapshots of particle positions and velocities, plus pre-computed power spectra at select snapshots; Fourier modes of the coarse-gridded density field; and the halo and subhalo catalogs, including how to index the halo catalogs and retrieve IDs of particles in halos.
    \item \textit{database\_examples}: How to query the halo database tables, including sample queries that demonstrate how to select from one run and snapshot, one run and multiple snapshots, and one snapshot and multiple runs.
    \item \textit{density\_field\_examples}: How to compute real-space density fields from the Fourier-space density fields and from the snapshots of particle positions, as well as how to create quick slices for plots using the Shape3D functionality.
    \item \textit{Shape3D\_examples}: How to use Shape3D objects to efficiently read subsets of particles contained in spheres, boxes, cones, and cone segments to e.g. grab all particles around (or in) a given halo or create lightcones.
\end{itemize}

\begin{figure}
	\includegraphics[width=\columnwidth]{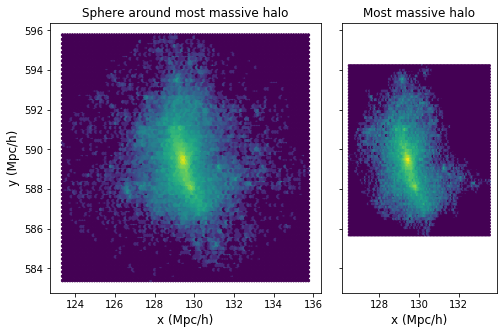}
    \caption{\textit{Left:} Particles in a sphere around the most massive halo in run 2\_0\_0 at $z=0$ whose radius is 3 times the radius of the halo. \textit{Right:} Particles identified as being in this halo.}
    \label{fig:Shape3D}
\end{figure}

Since reading all $1024^3$ particles from one snapshot can take some time, the Shape3D functionality can significantly speed up analyses that are only concerned with portions of the full volume. This functionality uses the spatial indexing library developed for numerical simulations~\citep{LemsonDBLP} to define shapes within a periodic volume and selectively read only those particles contained within the shape. Figure~\ref{fig:Shape3D} demonstrates the result of querying the most massive halo for run 2\_0\_0 at $z=0$ from the database and reading all particles in a sphere around that halo, which takes a few seconds to execute. The right panel shows only those particles identified as being within the halo, which can take 1-2 minutes because it requires reading the entire binary halo catalog and the halo particle IDs for that run and snapshot. This is one of the examples in the \textit{Shape3D\_examples} notebook.


\bsp	
\label{lastpage}
\end{document}